\documentstyle[12pt]{article}
\def\bmr{{\bf r}}
\def\bmx{{\bf x}}
\def\bmy{{\bf y}}
\def\bmz{{\bf z}}
\def\minus{\mbox{$-$}}

\begin{document}
\hspace{\fill} \fbox{\bf  LA-UR-97-1892}
\begin{center}

{\Large {\bf Higher-Order Nuclear-Size Corrections in Atomic Hydrogen}}\\

\vspace*{0.80in}

J.L.\ Friar \\
Theoretical Division \\
Los Alamos National Laboratory \\
Los Alamos, NM  87545 \\

\vspace*{0.1in}
and

Institute for Nuclear Theory\\
University of Washington\\
Seattle, WA 98195-1550\\

\vspace*{0.20in}
and
\vspace*{0.20in}

G.\ L.\ Payne\\
Department of Physics and Astronomy\\
University of Iowa\\
Iowa City, IA 52242\\

\end{center}

\vspace*{0.50in}

\begin{abstract}
Nuclear-size corrections of order $(Z \alpha )^5$ and $(Z \alpha )^6$ to the
S-state levels of hydrogenic atoms are considered. These nuclear-elastic
contributions are somewhat smaller than the polarizability (nuclear-inelastic)
corrections for deuterium, but are of comparable or larger size for the 
hydrogen case. For deuterium the (attractive) nonrelativistic $(Z \alpha )^5$
correction to the 2S-1S transition is 0.49 kHz, while the (repulsive) 
relativistic $(Z \alpha )^6$ contribution is \minus3.40 kHz. For the proton 
the corresponding corrections are 0.03 kHz and \minus0.61 kHz, respectively. 
The $(Z \alpha )^5$ contribution largely cancels the Coulomb-retardation part 
of the nuclear-polarization correction.

\end{abstract}
\pagebreak

\begin{center}
{\large {\bf Introduction}}\\
\end{center}

Recent experiments\cite{1,2,3} have pushed the precision of the spectroscopy of
the hydrogen isotopes to the point where previously unconsidered higher-order
terms (in $ \alpha$, the fine-structure constant) are now required. In the
isotope shift of the 2S-1S transition between deuterium and normal hydrogen, 
for example, experimental uncertainties\cite{4} of less than 0.2 kHz (one part
in $3 \cdot 10^4$ of the total nuclear-finite-size correction) have been
reported. The precision is such that these measurements afford us an
unparalleled opportunity to extract a precise value of the d-p
mean-square-radius difference. At the reported level of experimental precision,
this quantity is sensitive to ``exotic'' contributions to the charge
density\cite{5}, such as relativistic corrections and meson-exchange currents.
It therefore behooves us to calculate all higher-order contributions to the
(atomic) frequency shift of this size.

We report here a calculation of nuclear finite-size (i.e., nuclear-elastic)
corrections of orders ($Z \alpha )^5$ and ($Z \alpha )^6$, which supplement the
usual leading-order ($Z \alpha )^4$ term. Recently, the polarization
(nuclear-inelastic) corrections of order ($Z \alpha )^5$ were calculated for
deuterium\cite{6} by expanding that quantity in a series in $\bar{E} R \sim
\frac{1}{20}$, where $\bar{E}$ is the average (virtual) nuclear-excitation
energy and $R$ is a typical nuclear size. The first three orders in this series
are expected to be accurate at the level of roughly 0.01 kHz, although
individual nuclear observables in that expression cannot be determined to that
accuracy\cite{7}. Very recently, the leading-logarithm contributions to the
proton-polarizability correction were calculated\cite{8}, as well.

Fortunately, little analytic work is required to obtain the finite-size
corrections, since they were calculated many years ago in the context of muonic
atoms\cite{9,10}. The first three orders of corrections for the 
$n\underline{\rm th}$ S-state can be written in the form:
$$
\Delta E_n = \frac{2 \pi}{3} Z \alpha\, | \phi_n (0)|^2 
\left( \langle r^2 \rangle -\frac{Z \alpha \mu}{2} \langle r^3 \rangle_{(2)}
+ (Z \alpha)^2 F_{\rm REL} + (Z \alpha \mu)^2 F_{\rm NR} + \cdots \right)\, ,
\eqno (1)
$$
where $Z$ is the nuclear charge, $\langle r^m \rangle$ is the  $m\underline{\rm
th}$ moment of the nuclear charge distribution (normalized to unit charge),
$\mu$ is the reduced mass, $\phi_n (0)$ is the electron wave function at the
origin, and the Zemach moment\cite{11,9} $\langle r^3 \rangle_{(2)}$ is defined
by
$$
\langle r^m \rangle_{(2)} = \int d^3 r\, r^m\, \rho_{(2)} (r)\, , \eqno (2)
$$
where the convoluted (Zemach) charge density is given by
$$
\rho_{(2)} (r) = \int d^3 z\, \rho (|\bmz - \bmr |)\, \rho (z) \equiv 
\rho \otimes \rho \, . \eqno (3)
$$
The nonrelativistic correction $F_{\rm NR}$ (of relative order $(Z \alpha )^2
m^2_e R^2$, where $m_e$ is the electron mass) is negligible and will not be
considered further, while the corresponding relativistic correction is defined 
by
$$
F_{\rm REL} = -\langle r^2 \rangle \left( \langle \ln ( \beta r) \rangle
+ \left[ \psi (n) + \gamma - \frac{(5 n + 9)(n-1)}{4 n^2} \right] + \gamma
-2 \right) + I_{\rm REL}\, , \eqno (4)
$$
where $\psi(n)$ is the digamma function, $\gamma$ is Euler's constant, 
$\beta=2 Z \alpha \mu / n$, and
\begin{eqnarray*}
\hspace*{0.05in}I_{\rm REL} &=& -\frac{\langle r^3 \rangle \langle 1/r 
\rangle}{3}\\
&+& \int d^3 s\, \rho (s) \int d^3 t\, \rho (t)\; \Theta ( s-t ) 
\left[ (t^2 + s^2) \ln (t/s)
-\frac{t^3}{3 s} +\frac{s^3}{3 t} + \frac{s^2-t^2}{3} \right]\\
&+& 6 \int d^3 u\, \rho (u) \int d^3 t\, \rho (t) \int d^3 s\, \rho (s)\; 
\Theta ( u-t)\; \Theta ( t-s ) 
\left[ \frac{ s^2}{3} \ln (t/s) -\frac{s^4}{45 t u} \right. \\
&+& \frac{s^3}{9}
(\frac{1}{t} + \frac{1}{u}) 
+\left. \frac{s^2 t^2}{36 u^2} -\frac{2 s^2 t}{9 u}
+\frac{s^2}{9} \right]\, .  \hspace*{2.6in} (5)
\end{eqnarray*}
The n-dependent terms in Eq. (4) were calculated independently by 
Karshenboim\cite{12}. 

We first treat $F_{\rm REL}$ before discussing the smaller, nonrelativistic $(Z
\alpha )^5$ contribution. We calculate $\langle r^2 \rangle$, $\langle \ln (2
\alpha \mu r) \rangle$, and $I_{\rm REL}$ using the same techniques adopted
earlier to treat the nuclear-polarization observables\cite{6}. The proton 
charge distribution is taken to be an exponential (dipole form factor) with a
radius\cite{13} of 0.862~fm. All integrals are known analytically for this
case\cite{9}. The deuteron charge distribution is obtained by first solving for
the deuteron wave function using a variety\cite{14,15,16,17,18,19,20} of
first-generation (i.e., older) and second-generation\cite{21,22,23} (i.e.,
recent) potential models. The latter fit the nucleon-nucleon scattering data
very well; the best of them can be considered as alternative phase-shift
analyses. This ``bare-deuteron'' density is folded with the sum of proton and
neutron densities, and a spline fit is performed on the result. The neutron
density is taken to be that of a dipole form factor multiplied by $\lambda q^2$,
with $\lambda$ adjusted to fit the observed\cite{24} neutron mean-square charge
radius: \minus0.114 fm$^2$. Finally, double and triple integrals are performed
using the spline-fitted folded density (viz., the $\rho$'s in Eq.\ (5)).

\begin{table}[htb]
\centering

\caption{Deuteron finite-size corrections of order $(Z \alpha )^6$ for various
potential models. The mean-square radius of each potential model is $\langle 
r^2 \rangle$, the logarithmic radius is defined by $\langle \ln (2 \alpha \mu  
r)\rangle$, the relativistic density correlation is labelled $I_{\rm REL}$, and
the corresponding deuteron 2S-1S finite-size frequency shift is $\nu^{(6)}_{\rm
FS}$. The corresponding proton case is considered last using an exponential
charge distribution. Both shifts are repulsive.}
\hspace{0.25in}

\begin{tabular}{|l|cccc|} \hline

\multicolumn{1}{|c|}{\rule{0in}{3ex} Potential Model}&
\multicolumn{1}{c}{$\langle r^2 \rangle\, ({\rm fm}^{2})$}&
\multicolumn{1}{c}{$\langle \ln (2 \alpha \mu r )\rangle$}&
\multicolumn{1}{c}{$I_{\rm REL}\,({\rm fm^2})$}&
\multicolumn{1}{c|} {$\nu^{(6)}_{\rm FS}\, ({\rm kHz})$}\\[0.5ex] \hline \hline
\multicolumn{5}{|c|}{Second-Generation Potentials}\\ \hline

Argonne V$_{18} $ \rule{0in}{2.5ex} &4.507  & -9.770  &   -3.091 &   -3.41 \\
Reid Soft Core (93)   &4.505  & -9.771  &   -3.095 &   -3.41 \\
Nijmegen (loc-rel)    &4.507  & -9.771  &   -3.091 &   -3.41 \\
Nijmegen (loc-nr)     &4.498  & -9.773  &   -3.090 &   -3.41 \\
Nijmegen (nl-rel)     &4.496  & -9.773  &   -3.096 &   -3.40 \\
Nijmegen (nl-nr)      &4.494  & -9.776  &   -3.094 &   -3.40 \\
Nijmegen (full-rel)   &4.483  & -9.771  &   -3.098 &   -3.39 \\ \hline  

\multicolumn{5}{|c|}{First-Generation Potentials} \\ \hline

Bonn (CS) \rule{0in}{2.5ex} & 4.505 & -9.772 & -3.099 & -3.41 \\
Argonne V$_{14}$         & 4.556 & -9.762 & -3.108 & -3.45 \\
Nijmegen (78)            & 4.579 & -9.757 & -3.111 & -3.46 \\
Super Soft Core (C)      & 4.595 & -9.755 & -3.116 & -3.48 \\ 
de Tourreil-Rouben-Sprung& 4.530 & -9.766 & -3.098 & -3.43 \\
Paris                    & 4.516 & -9.768 & -3.094 & -3.42 \\
Reid Soft Core (68)      & 4.459 & -9.774 & -3.056 & -3.38 \\ \hline 

\multicolumn{5}{|c|}{Proton} \\ \hline

proton \rule{0in}{2.5ex} & 0.743 & -10.652 & -0.473 & -0.61 \\ \hline

\end{tabular}
\end{table}

The results are given in Table 1. We summarize the second-generation results 
for the deuteron as
\begin{eqnarray*}
\hspace*{2.05in} I_{\rm REL} &=& -3.094(4)\, {\rm fm}^2\\
\langle \ln (2 \alpha \mu r) \rangle &=& -9.773(3) \hspace*{2.3in} (6)\\
\nu_{\rm FS}^{(6)} &=& \minus3.40(1) \, {\rm kHz} \, .
\end{eqnarray*}
We caution that these ``uncertainties'' are merely spreads in the
potential-model results. Changing various aspects of the models that we have
used (including nucleon radii) could produce larger changes than these
uncertainties.

As an example of this caveat we note that the small Darwin-Foldy relativistic
correction to the proton and deuteron charge densities have not been included 
in our model. This is easily done by adding 0.0332 fm$^2$ to the $\langle r^2
\rangle$ of the proton (and hence to the deuteron). Consequently, the effect
cancels\cite{5} in the isotopic difference. This addition modifies the proton
charge radius to 0.881 fm, which increases $\nu^{(6)}_{\rm FS}$ by 0.03 kHz in
{\bf both} cases, largely by increasing $\langle r^2 \rangle$ rather than by
changing $I_{\rm REL}$ or $\langle \ln (2 \alpha \mu r) \rangle$.

Finally, we consider the $(Z \alpha )^5$ (second) term in Eq.\ (1). This is 
very similar in structure to the Coulomb-retardation part of the
nuclear-polarization correction\cite{6}, which can be written in the form
$$
\Delta E_n = - \frac{\pi}{3} \alpha^2\, m_e | \phi_n (0)|^2 
\left[ \int d^3 x \int d^3 y\, |\bmx - \bmy|^3\, \langle 0 | \rho^{\dagger} 
(\bmx ) \rho (\bmy) | 0 \rangle - Z^2 \langle r^3 \rangle_{(2)} \right]\, ,
\eqno (7)
$$
where the functions $\rho (\bmx )$ and $\rho (\bmy )$ in the correlation
function $\langle 0 | \rho^{\dagger} (\bmx ) \rho (\bmy) | 0 \rangle$ are
nuclear charge operators. Ignoring the difference between $m_e$ and $\mu$ 
(which is of recoil order), the last term in Eq.\ (7) exactly cancels the 
second term in Eq.\ (1), leaving only the correlation function. That is, the
separation of the nuclear Compton amplitude (which determines many of the
nuclear-structure-dependent atomic corrections) into nuclear-elastic and
nuclear-inelastic parts is somewhat artificial for this particular term.

The correlation function in Eq.\ (7) vanishes for certain (special) cases of
interest to us. For the deuteron case the correlation function generates two
small terms from the finite sizes of the proton (2.44 fm$^3 \rightarrow$ 0.032
kHz) and the neutron (\minus1.58 fm$^3 \rightarrow$ \minus0.020~kHz), which
largely cancel and leave a very small residue: 0.012~kHz. Both terms vanish in
the point-nucleon limit because $(\sum_i e_i)^2 \equiv \sum_i e^2_i$, where
$e_i$ counts the charge of the $i$\underline{\rm th} nucleon. We note that the
same cancellation takes place if we evaluate the proton correlation function in
the naive, nonrelativistic, pointlike quark model (where {\it mutatis mutandis}
$e_i$ counts the quark charges). In this case the individual cancelling
quantities are only 0.03~kHz in size.

\begin{table}[htb]
\centering
\caption{Contributions in kHz to the higher-order deuteron and proton
finite-size frequency shifts for the 2S-1S transition, together with their 
order in $(Z \alpha )$, differences between the deuteron and proton, and grand
totals. Negative contributions are repulsive. For comparison the
Coulomb-retardation nuclear-polarization correction, $\nu^{\rm ret}_{\rm pol}$,
is listed also, but is not included in totals.}

\hspace{0.25in}

\begin{tabular}{|l||c||cc|c|}
\hline
{Order}\rule{0in}{2.5ex} &$\nu^{\rm ret}_{\rm pol}$&$(Z\alpha)^5$ & 
$(Z \alpha )^6$ & Total\\ \hline \hline
{deuteron}\rule{0in}{2.5ex} & -0.48   & 0.49 & -3.40 &  -2.91 \\ 
{proton}  & -0.03   & 0.03 & -0.61 &  -0.58 \\ \hline
{Isotope Shift}&{}\rule{0in}{2.5ex} & 0.46 & -2.79 &  -2.33 \\ \hline

\end{tabular}
\end{table}

These results are displayed in Table 2, where the finite-size Coulomb
corrections are listed on the right, while the Coulomb-retardation 
contributions to the nuclear-polarization correction are listed (for comparison
only) on the left (they are not included in the totals). The cancellations are
evident.

In summary, we have computed the Coulomb nuclear-finite-size corrections of
orders $(Z \alpha )^5$ and $(Z \alpha )^6$ to the energy levels of hydrogenic
atoms. The relatively small $(Z \alpha )^5$ term largely cancels the
previously-calculated Coulomb-retardation term to the nuclear-polarization
corrections for the deuteron and proton. The $(Z \alpha )^6$ contribution is
larger because of a very large logarithm and because its relativistic origin
precludes the need for the very small nonrelativistic factors of ($m_e R \sim
\frac{1}{200}$). The net higher-order Coulomb nuclear-finite-size corrections
are \minus2.91~kHz for deuterium and \minus0.58~kHz for the proton.

\noindent {\bf Acknowledgements}

The work of J.\ L.\ Friar was performed under the auspices of the United States
Department of Energy, while that of G.\ L.\ Payne was supported in part by the
United States Department of Energy.

\end{document}